\documentstyle[eqsecnum,aps]{revtex}

\begin{document}
\title{The astrophysical reaction $^8Li(n,\gamma )^9Li$ from measurements by
reverse kinematics}
\author{Carlos A. Bertulani}
\address{$^a$ Instituto de F\'\i sica, Universidade Federal
do Rio de Janeiro\\
21945-970 \ Rio de Janeiro, RJ, Brazil. E-mail:
bertu@if.ufrj.br}
\date{\today }
\maketitle

\begin{abstract}
We study the breakup of $^9Li$ projectiles in high energy (28.5 MeV/u)
collisions with heavy nuclear targets ($^{208}Pb$). The wave functions are
calculated using a single-particle model for $^9Li$, and a simple optical
potential model for the scattering part. A good agreement with measured data
is obtained with insignificant E2 contribution.
\end{abstract}

\bigskip

\noindent


The use of the Coulomb dissociation method \cite{Ba86,Be88} has proven to be
a useful tool for extracting radiative capture reaction cross section of
relevance for nuclear astrophysics. In particular it appears that the
Coulomb dissociation of $^9Li$ is very useful \cite{Ze98} for elucidating
the role of the inhomogeneous nucleosynthesis in the Big Bang model - the
formation of $^9Li$ via the $^8Li(n,\gamma )^9Li$ reaction. However, a few
lingering questions still need to be addressed, including the importance of
E2 excitations for the kinematics of the MSU experiment \cite{Ze98},
performed at approximately 28.5 MeV/u. We attempt to resolve this issue by
using a relatively simple but still realistic nuclear model that however
yields a very good agreement with data and suggest that E2 excitations are
negligible for the kinematical conditions of the MSU experiment.

For $^9Li$ assume that the $J_0=3/2^{-}$ ground state can be described as a $%
j_0=p_{3/2}$ proton coupled to the $I_c=2^{+}$ ground state of the $^8Li$
core. The spectroscopic factor for this configuration was taken as unity.
The single particle states, $R_{E_xlj}(r)$, for the excitation energy $%
E_x=E_n+\left| E_0\right| $ ( = 4.05 MeV), where $E_n$ is the neutron-$^8Li$
relative energy, are found by solving the Schr\"{o}dinger equation with a
nuclear potential with spin-orbit of the form

\begin{equation}
V(r)=V_0\left[ 1-F_{s.o.}\left( {\bf l.s}\right) \frac{r_0}r\frac d{dr}%
\right] f\left( r\right) \;,\;\;\;\;\;\;\;\;f\left( r\right) =\left[ 1+\exp
\left( \frac{r-R}a\right) \right] ^{-1}\;,
\end{equation}
with the parameters $a=0.52$ fm, $r_0=1.25$ fm, $R=2.5$ fm, $F_{s.o.}=0.351$
fm, and $V_0=-45.3$ MeV. This reproduces the ground state bound energy $%
E_0=-4.05$ MeV. The same set of parameters was used to calculate the
continuum wavefunctions with energy $E_x$. The total wavefunction is
constructed by coupling the single-particle states with the spin of $^8Li $.
To compute the S-factors for the capture process $b+c\rightarrow a$ we have
used the first-order perturbation theory and the reduced matrix elements for
electric multipole transitions given by

\begin{eqnarray}
\left\langle j\left\| {\cal M}(E{\lambda })\right\| j_0\right\rangle
&=&e_\lambda \left( -1\right) ^{l_0+l+j_0-j}\;\left[ \frac{\left( 2\lambda
+1\right) \left( 2j_0+1\right) }{4\pi \left( 2j+1\right) }\right]
^{1/2}\left\langle j_0\lambda \frac 120\mid j\frac 12\right\rangle  \nonumber
\\
&&\times \left[ \frac{1+\left( -1\right) ^{l_0+l+\lambda }}2\right] \;\int
R_{E_xlj}(r)\;R_{E_0l_0j_0}(r)\;r^{\lambda +2}\;dr
\end{eqnarray}
where $e_\lambda =Z_be(-A_x/A_a)^\lambda +Z_xe(A_b/A_a)^\lambda $ is the
effective electric charge. We have considered $s,p,d$ and $f$ continuum
states.

The response functions for the excitation of $^9Li$ are defined by ($b\equiv
\;^8Li$, $c\equiv n$)

\begin{equation}
\frac{dB\left( E\lambda ;\;l_0j_0\rightarrow E_xlj\right) }{dE_x}=\frac{%
m_{bc}}{\hbar ^2k}\;\left| \left\langle j\left\| {\cal M}(E{\lambda }%
)\right\| j_0\right\rangle \right| ^2/\left( 2j_0+1\right) \;
\end{equation}
They are presented in figure 1 for $E1$ and $E2$ excitations, as functions
of the neutron energy, $E_n=\hbar^2/2m_{bc}$. They are dominated by s-wave
components: the higher angular momentum waves become relevant for higher
energies, as expected.

The cross sections for direct capture are given by

\begin{equation}
\sigma _{DC}^{\left( \lambda \right) }\left( E_x\right) =\frac{\left(
2I_c+1\right) }{\left( 2J_0+1\right) }\;\frac{\left( 2\pi \right) ^3\left(
\lambda +1\right) }{\lambda \left[ \left( 2\lambda +1\right) !!\right] ^2}\;%
\frac 1{k^2}\;\left( \frac{E_x}{\hbar c}\right) ^{2\lambda +1}\frac{dB\left(
E\lambda \right) }{dE_x}\;.
\end{equation}

The figure 2 displays the direct capture cross section for the $E1$
transitions. The experimental data are from ref. \cite{Ze98}. Due to the
factor $\left( E_x/\hbar c\right) ^{2\lambda +1}$ appearing in eq. (4) (and
since $B(E1)/e^2$ and $B(E2)/e^4$ are of same order of magnitude, as seen
from fig. 1), the $E2$ contribution to the radiative capture cross section
is about 7 orders of magnitude smaller than the $E1$ contribution. For
pedagogical purposes, we also show in figure 2 the hard sphere model of Lane
and Lynn \cite{LL60} for the direct capture cross section. The hard sphere
model is the simplest model to calculate the reaction cross section for $%
\left( n,\gamma \right) $-processes at low energies. This model has been
used by other authors with reasonable success in some situations \cite
{LL60,Mug82,MC91}. According to this model, the direct capture cross section
for the $n+A\;({\rm target})\longrightarrow \left( A+1\right) +\gamma $
reaction is given by

\begin{equation}
\sigma _\gamma ^{(d.c.)}=\sigma _{h.s.}\left[ 1+\frac{R-a_s}R\frac{y\left(
y+2)\right) }{y+3}\right] ^2
\end{equation}
where $R$ is the target radius, which we take as $R=1.35\;A^{1/3}$ fm, $%
y^2=2\mu _{nA}E_xR^2/\hbar $, $\mu _{nA}$ being the neutron+$^8Li$ reduced
mass, and $A=8$. The scattering length for the $n+\;^8Li$ system at low
energies is $a_s=2.03$ fm \cite{KS77}.

The hard-sphere, $\sigma _{h.s.}$, cross section entering eq. (5) is given by

\begin{equation}
\sigma _{h.s.}=\frac{0.062}{R\sqrt{E_n}}\left( \frac ZA\right) ^2\frac{2J_0+1%
}{6(2I_c+1)}\cdot S\cdot \left[ \frac{y\left( y+3\right) ^2}{y+1}\right] ^2\;%
{\rm mb\;,}
\end{equation}
with the neutron energy, $E_n$, given in keV. In this equation, $Z$ is the $%
^8Li$ charge, $I_c$ its spin, $J_0$ the spin of $^9Li$, and $S$ is the
spectroscopic factor, which we take as unity. The hard sphere model yields a
cross within only a factor of 2 difference from our results for E1
transitions, but with approximately the same energy dependence.

Since there is no data for the elastic scattering of $^9Li$ on $Pb$ targets
at this bombarding energy, we construct an optical potential using an
effective interaction of the M3Y type \cite{KBS84,BLS97} modified so as to
reproduce the energy dependence of total reaction cross sections, i.e. \cite
{BLS97}, 
\begin{equation}
t(E,s)=-i{\frac{\hbar v}{2t_0}}\;\sigma _{NN}(E)\;\left[ 1-i\alpha
(E)\right] \;t(s)\ ,  \label{tes}
\end{equation}
where $t_0=421$ MeV is the volume integral of the M3Y interaction $t(s)$, $v$
is the projectile velocity, $\sigma _{NN}$ is the nucleon-nucleon cross
section, and $\alpha $ is the real-to-imaginary ratio of the forward
nucleon-nucleon scattering amplitude. At 28.5 MeV/nuc, we use $\sigma
_{NN}=20$ fm$^2$ and $\alpha =0.87$.

The optical potential is given by 
\begin{equation}
U(E,{\bf R})=\int d^3r_1\;d^3r_2\;\rho _{_P}({\bf r}_1)\rho _{_T}({\bf r}%
_2)\;t(E,s)\ ,
\end{equation}
where ${\bf s}={\bf R}+{\bf r}_2-{\bf r}_1$, and $\rho _{_T}$ ($\rho _{_P}$)
is the ground state density of the target (projectile). 

Following ref. \cite{BN93}, the Coulomb amplitude is given by 
\begin{equation}
f_C=\sum_{\lambda \mu }f_{\lambda \mu }^{(JM)}\ ,
\end{equation}
where 
\begin{eqnarray}
f_{\lambda \mu }^{(JM)} &=&i^{1+\mu }\;{\frac{Z_Te\mu _{_{PT}}}{\hbar ^2}}%
\;\left( {\frac{E_x}{\hbar c}}\right) ^\lambda \;\sqrt{2\lambda +1}\;\exp
\left\{ -i\mu \phi \right\} \;\Omega _\mu (q)  \nonumber \\
&&\times G_{E\lambda \mu }({\frac cv})\left\langle JM\left| {\cal M}%
_{E\lambda ,-\mu }\right| J_0M_0\right\rangle \ ,
\end{eqnarray}
\begin{equation}
\Omega _\mu (q)=\int_0^\infty db\;b\;J_\mu (qb)K_\mu \left( {\frac{E_xb}{%
\gamma \hbar v}}\right) \exp {i\chi (b)}\ .
\end{equation}
$J_\mu (K_\mu )$ is the cylindrical (modified) Bessel function of order $\mu 
$, and the functions $G_{\pi \lambda \mu }(c/v)$ are tabulated in ref. \cite
{AW79}. The angular momentum algebra connects $\left\langle JM\left| {\cal M}%
_{E\lambda ,-\mu }\right| J_0M_0\right\rangle $ with the reduced matrix
elements of eq. (2).

The eikonal phase, $\chi (b)$, is given by 
\begin{equation}
\chi (b)=2\eta \ln (kb)-{\frac 1{\hbar v}}\;\int_{-\infty }^\infty
dz\;U_{opt}(R)\ ,
\end{equation}
where $\eta =Z_PZ_Te^2/\hbar v$, $k$ is the projectile momentum, and $R=%
\sqrt{b^2+z^2}$. The optical potential, $U_{opt}$, in the above equation is
given by eq. (8).

The cross section for Coulomb excitation to the state with angular momentum $%
J$ and excitation energy $E_x$ is obtained by an average (and a sum) over
the initial (final) angular momentum projections: 
\begin{equation}
\frac{d\sigma _\lambda }{d\Omega dE_x}={\frac 1{2J_0+1}}\;\sum_{M_0,\
M}\left| f_{\lambda \mu }^{(JM)}\right| ^2\;.
\end{equation}

As explained in details in refs. \cite{Ba86,Be88,BN93}, the above cross
section can be factorized in terms of a product of  virtual photon numbers
and breakup cross sections by real photons, which by detailed balance are
directly related to the radiative capture cross sections. Thus a measurement
of $d\sigma _\lambda /d\Omega dE_x$ can be used to obtain radiative capture
cross sections of astrophysical interest. Experimentally, the nuclear
contribution to the breakup cross section can be separated by repeating the
measurement on light targets (see, e.g., ref. \cite{Ze98}). 

At the bombarding energies of tens of MeV/nucleon, the $E2$ virtual photon
number is much larger than that of $E1$. As a consequence, even when the $E2$
contribution to the radiative capture cross section is small, they are
amplified in Coulomb breakup experiments \cite{Be88}. In order to infer the
relevance of $E2$ for the breakup of $^9Li$ (28.5 MeV) on lead targets, we
plot in figure 3 the angle integrated cross section $d\sigma _\lambda /dE_x$%
, as a function of the neutron energy relative to $^9Li$. We see that the $E2
$ contribution to the breakup cross section is 3 orders of magnitude smaller
the $E1$ contribution. We thus conclude that $E2$ transitions are not
relevant in the experiment of ref. \cite{Ze98}. Although we obtained this
result by means of a single particle model for the $^8Li(n,\gamma )^9Li$
reaction, we do not expect that it would change appreciably with more
sophisticated models.

\bigskip\bigskip
\noindent{\bf Acknowledgments}

\medskip

I am grateful to Aaron Galonsky for helpful comments and suggestions. This
work was supported in part by MCT/FINEP/CNPQ(PRONEX) under contract No.
41.96.0886.00,and by the Brazilian funding agencies FAPERJ and FUJB.

\bigskip 
{\bf Figure Caption}\\

{\bf Fig. 1} - Response function in units of $e^2fm^2/MeV$ ($e^2fm^4/MeV$)
for $E1$ ($E2$) transitions in the reaction $^9Li(\gamma ,n)^8Li$, as a
function of the neutron energy relative to the $^8Li$ core. The lower curves
are the f (solid) and p+f (dashed) -waves contribution, while the upper
curves display the contribution of s (solid) and s+d (dashed) -waves.

{\bf Fig. 2} - Radiative capture cross sections, in $\mu b$, for the
reaction $^8Li(n,\gamma )^9Li$ in the direct capture model. The solid line
is obtained with the s-waves, while the dashed line includes transitions to
d-waves. The hard-sphere model of ref. \cite{LL60} is also shown
(dashed-dotted line). The experimental data are from ref. \cite{Ze98}.

{\bf Fig. 3} - Coulomb breakup cross section $d\sigma _\lambda /dE_n$ (in
mb/MeV) for the reaction $^9Li\;$(28.5 MeV/nucleon) $+Pb\longrightarrow
\;^8Li+n$, as a function of the neutron-$^9Li$ relative energy, in MeV.  The
E2 breakup contribution is multiplied by $10^3$ to be shown in the same
graph.

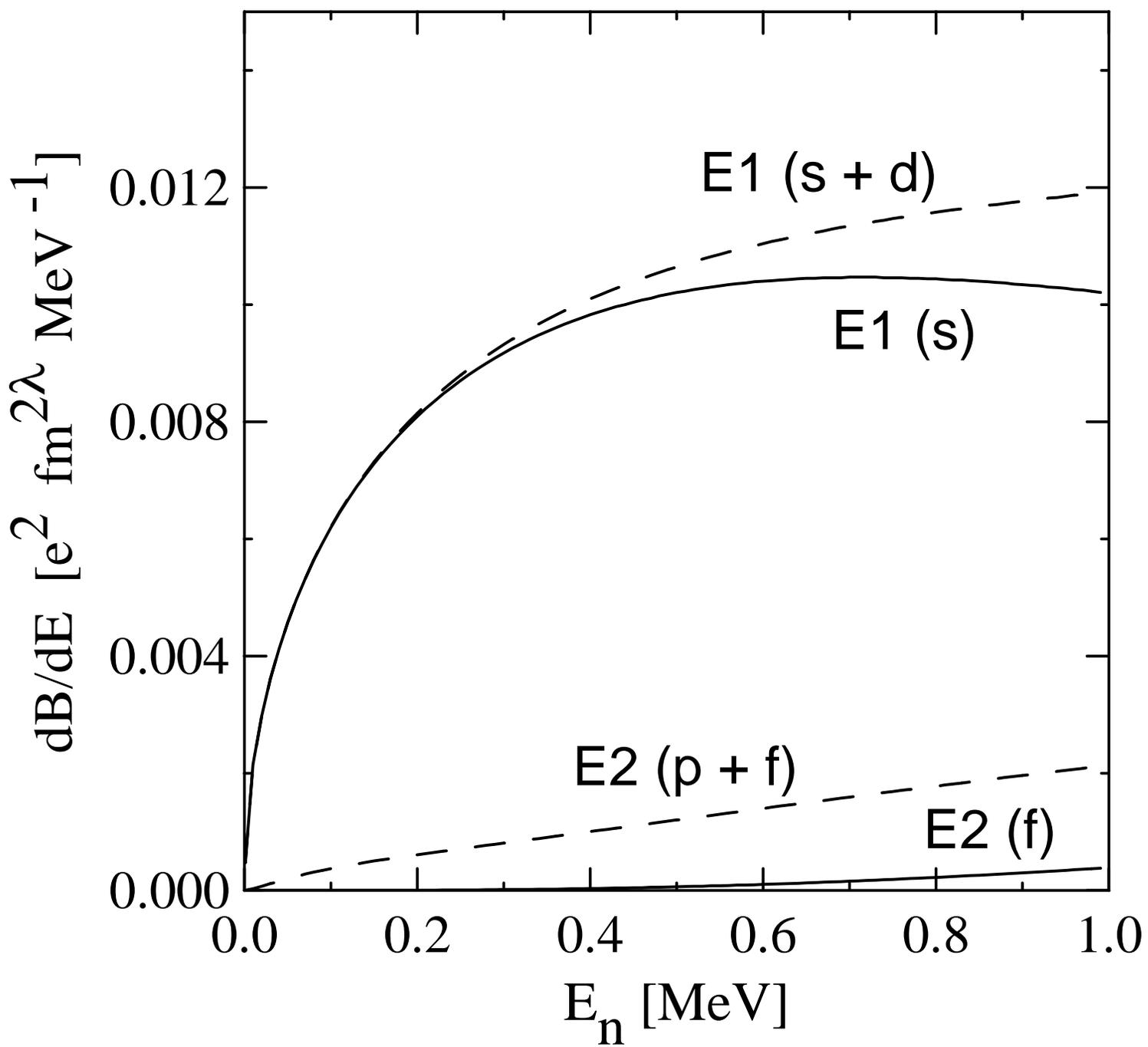
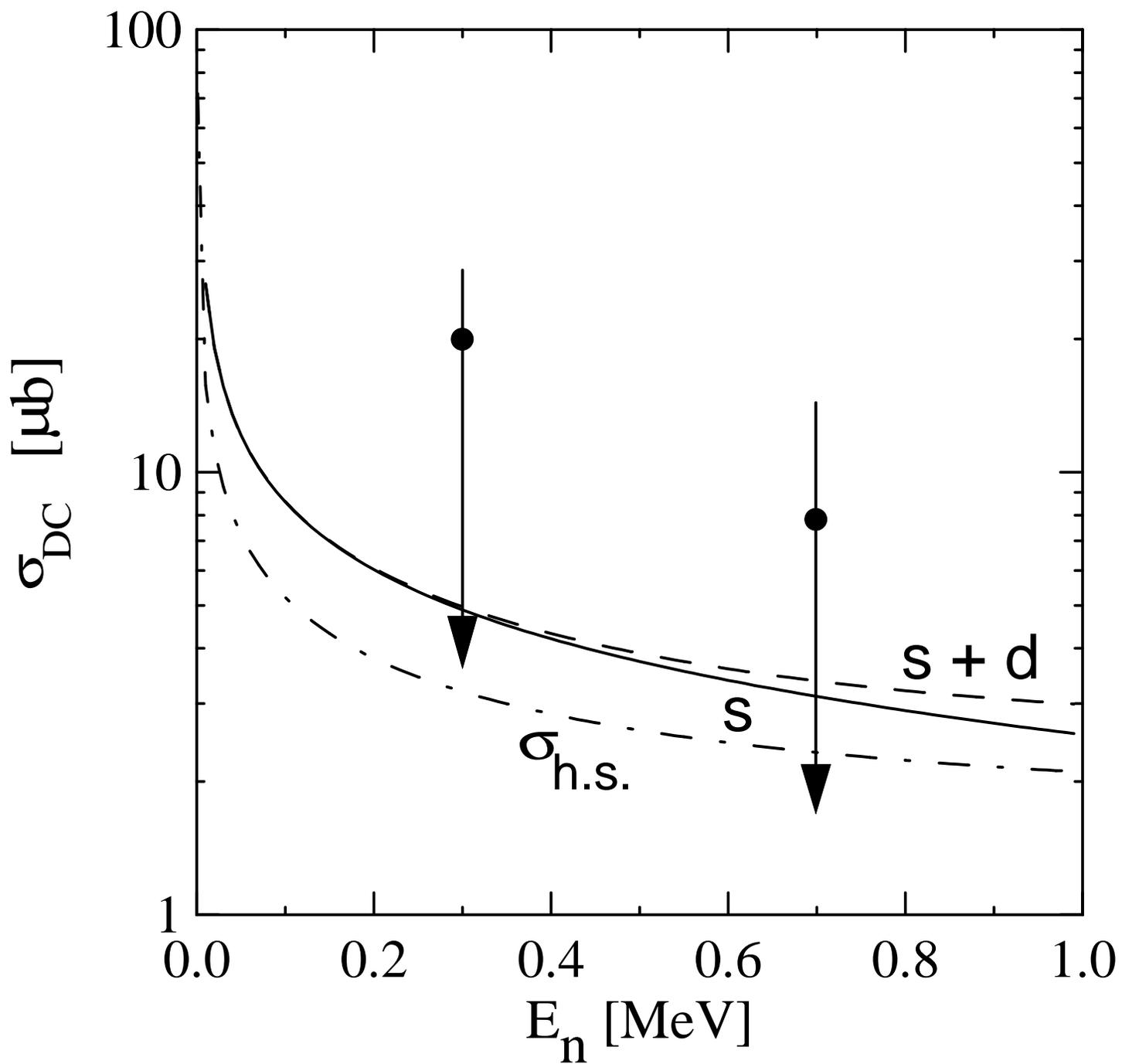
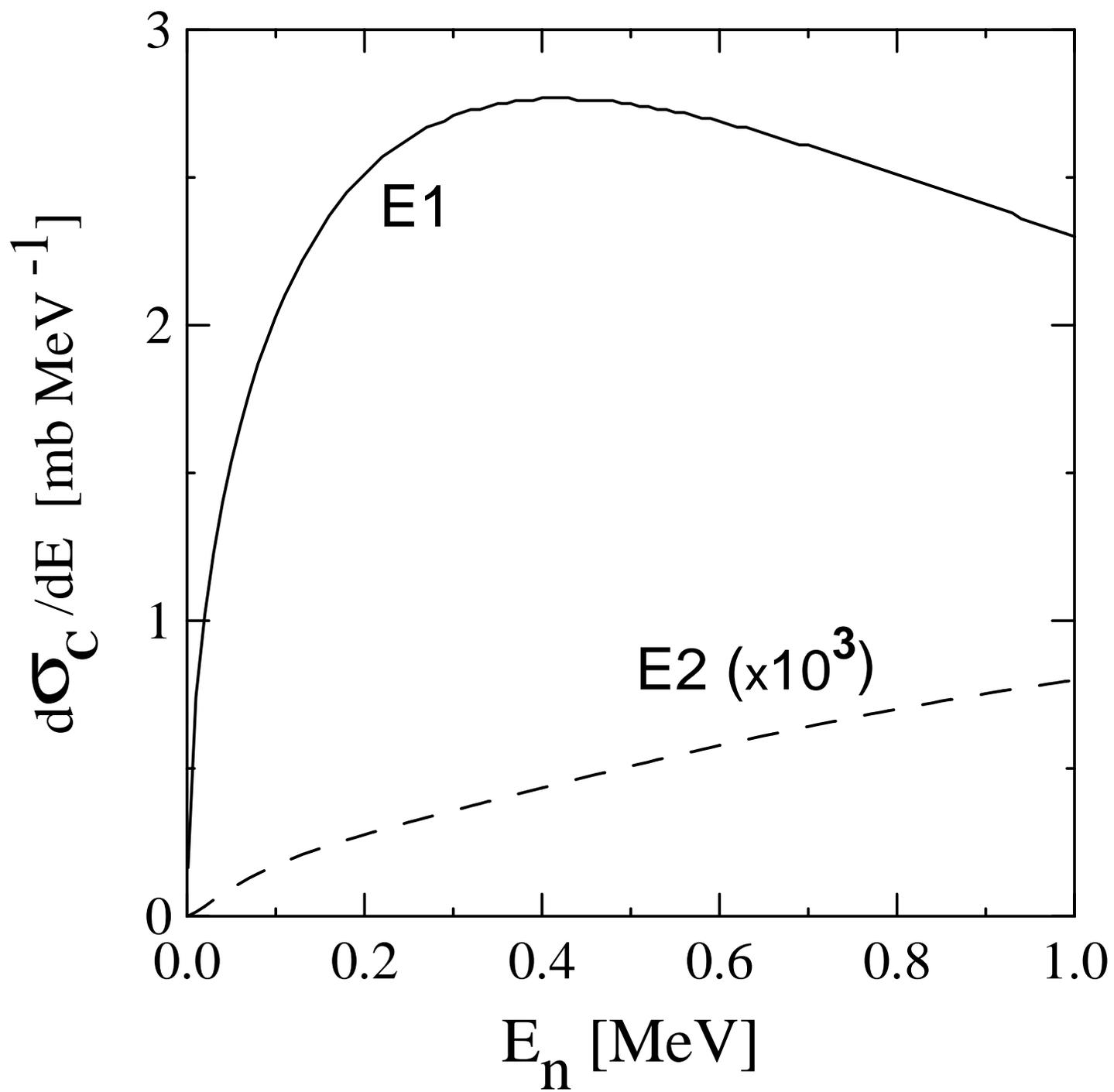
\end{document}